\documentclass[preprint,12pt,authoryear]{elsarticle}

\usepackage{amssymb,amsmath}
\usepackage{booktabs}
\usepackage{bm}
\usepackage{graphicx}
\usepackage{url}
\usepackage[usenames,dvipsnames]{color}
\definecolor{purple}{rgb}{0.5,0,0.5}

\newcommand{\Fleq}{\mathcal{F}_{\leq k}}
\newcommand{\rhok}{\rho(k)}
\newcommand{\dd}{{\rm d}}

\journal{Journal of Symbolic Computation}

\begin{document}

\begin{frontmatter}

\title{Exhaustive Symbolic Integration: Integration by Differentiation and the Landscape of Symbolic Integrability}

\author[icg]{Harry Desmond\corref{cor1}}
\cortext[cor1]{Corresponding author}
\ead{harry.desmond@port.ac.uk}

\affiliation[portsmouth]{organization={Institute of Cosmology and Gravitation},
  addressline={University of Portsmouth},
  city={Portsmouth}, postcode={PO1 3FX}, country={United Kingdom}}

\begin{abstract}
We introduce Exhaustive Symbolic Integration (ESI), a method that enumerates
all symbolic functions up to a given complexity~$k$ within a specified operator
basis and determines which admit closed-form antiderivatives within the same
class.  This allows us to compute the ``integrability fraction''
$\rhok$
(the fraction of functions whose derivatives lie within the same class), which we do for five operator bases including combinations of rational functions, powers, exponentials, logarithms and trigonometric functions. We find that $\rhok$ declines at high complexity and that the operator basis has a dramatic effect---in particular, adding the logarithm boosts $\rhok$ by a factor of $\sim$3 and produces or exacerbates a clear peak at $k=6$.
We also deploy ESI as a novel integration algorithm, identifying three integrals
that resist SymPy, Mathematica, RUBI, FriCAS, Maxima and Giac under all tested strategies.
When an antiderivative can be found by multiple methods, ESI often returns the simplest form.
These results reveal that the landscape of symbolic integrability is shaped primarily by the choice of operators, and that exhaustive enumeration can systematically discover integrable forms---including novel ones---that elude computer albegra systems.
\end{abstract}


\end{frontmatter}

\section{Introduction}
\label{sec:intro}

Differentiation is algorithmic: given any expression built from elementary
functions, its derivative can be computed mechanically by the chain and product
rules.  Integration admits no such universal algorithm \citep{Bronstein2005}.  The decision procedure
of \citet{Risch1969}, building on the structure theorem of \citet{Liouville1835},
determines whether a given elementary function possesses an elementary
antiderivative, and computes one when it exists.  Yet neither Liouville's
theorem nor the Risch algorithm addresses a more basic structural question:
within a given class of symbolic expressions, how \emph{prevalent} is
integrability?  That is, if one draws a function at random from a well-defined
symbolic grammar, what is the probability that its antiderivative can be
expressed within the same grammar?

We make this question precise as follows.  Fix an operator basis (a set of nullary, unary and binary operations) and let $\Fleq$ denote the set of all distinct functions
expressible as expression trees with at most $k$ nodes.  Differentiation maps each function $F \in \Fleq$ to a derivative $F'$ that may or may not itself belong to $\Fleq$.  The
\emph{integrability fraction}
\begin{equation}\label{eq:1}
	\rhok \;=\; \frac{|\{F \in \Fleq : F' \in \Fleq\}|}{|\Fleq|}	
\end{equation}
measures the degree to which $\Fleq$ is closed under the inverse of
differentiation.  Because $\Fleq$ is finite and explicitly constructible,
$\rhok$ is exactly computable for any chosen basis and complexity bound.

This quantity connects to classical differential algebra in a way that is both natural and novel.
The elementary functions---those built from rational functions, exponentials and logarithms---form a differential field: they are closed under the field operations and under differentiation. Liouville's theorem \citep{Liouville1835,Rosenlicht1972} characterises when an elementary function admits an elementary antiderivative; in many cases no such antiderivative exists, so the class is not closed under integration. This is a qualitative statement about integrability in the unrestricted field, and does not address what happens at finite complexity, where one restricts to expressions of bounded size. The integrability fraction $\rho_k$ provides the quantitative complement: it measures how often functions within such a bounded-complexity subset admit elementary antiderivatives, and how this depends on the operators from which the functions are built.
The sensitivity of $\rhok$ to the basis reveals which operations promote closure and which demote it.

To address these questions we introduce a novel method \emph{Exhaustive Symbolic Integration} (ESI), which computes antiderivatives by differentiating candidate solutions within exhaustive function sets. We study five operator bases exhaustively up to a basis-dependent maximum complexity $k=8$--$10$, comprising
over $10^6$ unique functions.
The bases range
from a minimal set of rational operations and powers through the
addition of exponentials and logarithms to an
independent trigonometric basis containing sine and cosine; crucially, the trigonometric basis is not nested within any of the others so provides a fully independent test of structural patterns.
ESI also affords an integration algorithm in its own right, locating antiderivatives by searching exhaustive catalogues for functions whose derivatives match the target integrands. This can be used both to find antiderivatives that fall through the cracks of computer algebra systems' Risch implementations and to rapidly produce simple antiderivative forms.

The idea of identifying integrals by differentiating expressions
has been explored in the machine learning literature \citep{LampleCharton2020}, and refined using Risch--Liouville theory to produce guaranteed-integrable expressions \citep{BarketEG2023,BarketEG2024}. These works generate expressions \emph{randomly} for the purpose of training neural networks, so do not measure systematic integrability fractions or study their dependence on the operator basis. CAS benchmarking suites \citep{Nasser2024} compare the performance of integration engines on curated test sets, but again do not address the prevalence of integrability in any defined function space. To our knowledge, the systematic, exhaustive measurement of $\rhok$ as a function of complexity and operator basis has not been attempted before,
nor has exhaustive enumeration been used to systematically discover novel closed-form integrals.

The paper is organised as follows.  Section~\ref{sec:method}
describes ESI, including the construction of the function spaces and differentiation and
fingerprinting pipeline.
Section~\ref{sec:results} presents the main results: the integrability
fraction, exploration of the logarithm's outsized effect, the structure
of the derivative map, novel antiderivatives and the comparison with CAS.
Section~\ref{sec:discussion} discusses limitations, implications for differential algebra
and directions for future work. Section~\ref{sec:conclusion} concludes.

\section{The Exhaustive Symbolic Integration method}
\label{sec:method}

\subsection{Function space construction and operator bases}
\label{sec:esr}

We construct the function space $\Fleq$ using Exhaustive Symbolic Regression \citep[ESR;][]{Bartlett2024}, which enumerates all expressions that can be built from a given operator basis up to a maximum complexity $k$.
While ESR was designed, and has been extensively used, for symbolic regression~\citep{Bartlett2024,RAR,Inflation,Martin_1,Martin_2,Kronberger2024,Ford}, here we utilise only its exhaustive function generation step. Complexity is defined as the number of nodes in the expression tree: each expression corresponds to a tree
in which internal nodes carry operators and leaves carry either the independent variable $x$ or free parameters $a_i$. For example, the expression $(a_0 + x)^2$ has complexity $k = 4$ (two leaves, one addition node and one squaring node, assuming all those operators are in the basis set), whereas the algebraically equivalent $x^2 + 2a_0 x + a_0^2$ has complexity $k = 11$, illustrating that complexity in this sense measures representational cost rather than mathematical sophistication.

At each complexity level, ESR applies symbolic simplification to remove duplicate expressions, so that $\Fleq$ contains only structurally distinct functions. This deduplication is essential: without it, the same mathematical function would appear many times in different syntactic forms, inflating the space and corrupting the integrability fraction.
ESR's deduplication proceeds in rounds: expressions are converted to SymPy objects, simplified (up to a 60-second timeout per expression), and compared symbolically.
As noted in \cite{Bartlett2024} and subsequent applications this is an imperfect procedure, with many duplicates remaining because SymPy simplification is incomplete. We therefore apply a second deduplication layer via numerical fingerprinting (Section~\ref{sec:fingerprint}) to catch additional duplicates.

We study five operator bases, ordered in Table~\ref{tab:bases} roughly by richness.
These bases probe how specific operators affect closure under the inverse of differentiation. \texttt{core\_maths} generates rational functions and powers, serving as a baseline. \texttt{core\_log\_maths} isolates the effect of adding \texttt{log}. \texttt{ext\_maths} adds \texttt{sqrt}, \texttt{sq}, and \texttt{exp}, capturing exponential growth. \texttt{ext\_log\_maths} further adds \texttt{log} (the exp--log pair has special closure properties explored in Section~\ref{sec:mechanism}), while \texttt{trig\_maths} replaces \texttt{sqrt}, \texttt{sq}, and \texttt{exp} with \texttt{sin} and \texttt{cos} and is not a superset of any other basis.
The spaces grow roughly exponentially with $k$; at the highest computed complexities they range from 77{,}053 unique \texttt{core\_maths} functions at $k=10$ to 1{,}454{,}666 unique \texttt{ext\_log\_maths} functions at $k=9$.
These counts define the denominator of~$\rhok$.  Table~\ref{tab:sizes} gives the cumulative function-space sizes and the CAS comparison counts used later. The \texttt{core\_log\_maths} basis was introduced to investigate the effect of log in the integrability-fraction analysis and is not included in the full six-engine CAS cascade, although we do report the available multi-member count and any lightweight CAS checks completed for it.

\begin{table}[t]
\centering
\caption{Operator bases studied.}
\label{tab:bases}
\begin{tabular}{llll}
\toprule
Basis & Nullary & Unary & Binary \\
\midrule
\texttt{core\_maths}    & $x$, $a$ & inv & $+, -, \times, \div$, pow \\
\texttt{core\_log\_maths} & $x$, $a$ & inv, log & $+, -, \times, \div$, pow \\
\texttt{ext\_maths}     & $x$, $a$ & inv, sqrt, sq, exp & $+, -, \times, \div$, pow \\
\texttt{ext\_log\_maths} & $x$, $a$ & inv, sqrt, sq, exp, log & $+, -, \times, \div$, pow \\
\texttt{trig\_maths}     & $x$, $a$ & inv, sin, cos & $+, -, \times, \div$, pow \\
\bottomrule
\end{tabular}
\end{table}

\begin{table}[t]
\centering
\caption{Deduplicated cumulative function counts and CAS integrability outcomes by basis.}
\label{tab:sizes}
\begin{tabular}{lrrcccc}
\toprule
 & & & \multicolumn{2}{c}{SymPy} & \multicolumn{2}{c}{Mathematica} \\
\cmidrule(lr){4-5} \cmidrule(lr){6-7}
Basis ($k_{\max}$) & $|\Fleq|$ & tested & hard & imposs. & hard & imposs. \\
\midrule
\texttt{core\_maths} (10)     & 77{,}053  & 13{,}993  & 1{,}059       & 218  & 2{,}478       & 411            \\
\texttt{core\_log\_maths} (8) & 21{,}214   & 5{,}669       & 144           & 59  & ---           & ---            \\
\texttt{ext\_maths} (9)       & 628{,}400  & 117{,}674 & 8{,}183       & 3{,}321 & 905           & ---            \\
\texttt{ext\_log\_maths} (8)  & 222{,}838  & 49{,}538  & 2{,}431       & ${\geq}232$   & 2{,}049       & 537  \\
\texttt{trig\_maths} (9)      & 897{,}293 & 51{,}027 & 9{,}795    & 3{,}700 & 685       & 435            \\
\bottomrule
\end{tabular}
\end{table}

Generating the function space is the most expensive step of the entire procedure: the cost grows rapidly with $k$ because the number of candidate trees is exponential. As an example, tree generation for \texttt{ext\_log\_maths} at $k=9$ requires ${\sim}20$ CPU-hours, and symbolic deduplication of the resulting 56 million candidate trees requires several days on tens of cores with ${\sim}20$--30\,GB of memory per core.
Thus for computational reasons we extend \texttt{core\_maths}, containing relatively few operators, to $k=10$ but the others only to $k=9$. We do not attempt a complete benchmark across all function sets; instead we report the entries that support the conclusions or are available from the completed pipeline.  The ``tested'' column is the number of multi-member equivalence classes, the only classes on which ESI supplies a non-trivial lookup primitive.  ``Hard'' denotes failure in the initial sweep,
while ``Imposs.'' denotes the longer stress tests where those were run.  The \texttt{ext\_log\_maths} SymPy entry is a lower bound because the stress test was applied to the 503 integrals failed by both SymPy and Mathematica, not to all 2{,}431 SymPy failures.

\subsection{Differentiation and numerical fingerprinting}
\label{sec:fingerprint}

After ESR's symbolic deduplication we canonicalise parameter names: if a function contains free parameters, we rename them $a_0, a_1, \ldots$ in the order they first appear when reading the expression tree of $F(x)$ from left to right. This ensures that e.g.\ $a_1 x$ and $a_0 x$ are recognised as the same parametric family. We then differentiate each $F(x)$ with respect to $x$ using SymPy \citep{Meurer2017}. Although integration and simplification can be challenging for SymPy, differentiation through the chain rule is straightforward. We then use a numerical fingerprinting scheme to remove surviving duplicates.

Each expression is evaluated at $N = 60$ random points. The independent variable $x$ is drawn uniformly from the interval $(0.2, 5)$,
and the parameters $a_i$ are assigned fixed values drawn uniformly from $(0.5, 3.0)$ using a deterministic random seed, ensuring reproducibility. All evaluations are performed using \texttt{mpmath} \citep{Meurer2017} at 50-digit precision. Each resulting value is then rounded to 10 significant figures, and the tuple of 60 rounded values is hashed via MD5 to produce a compact fingerprint. Two expressions with identical hashes are treated as functionally equivalent. Note that because all expressions are evaluated at the same fixed parameter values, this scheme identifies duplicates only when they agree pointwise for the same parameter assignments.

The reliability of this scheme rests on the analytic properties of the functions involved. If two expressions $f$ and $g$ are genuinely distinct, their difference $h \equiv f - g$ is a non-zero analytic function with only isolated zeros. The probability that $|h|$ falls below the rounding threshold at any single random evaluation point is at most ${\sim}10^{-10}$. Under random choice of the evaluation seed the expected per-pair false-positive probability over $N = 60$ points is therefore $\lesssim 10^{-600}$; for any single fixed seed, the bound is a typical-case expectation rather than a worst-case guarantee, since pathological functions with zeros at all 60 evaluation points would evade detection. This makes the false-positive rate negligible even in the largest datasets containing up to ${\sim}10^{10}$ distinct pairs.
False negatives---genuinely equivalent expressions assigned different hashes---can occur if evaluation fails at some points due to domain restrictions (e.g.\ logarithms of negative arguments, division by zero).

We exclude at the outset all expressions with no $x$-dependence (parameter-only expressions). After this cut, the current \texttt{trig\_maths} $k\leq 8$ run has 150{,}388 successfully fingerprinted functions, 78 explicit timeouts, and 354 additional fingerprinting failures.
These timeouts occur when \texttt{mpmath} evaluation hangs on deeply nested compositions (e.g.\ iterated exponentials that overflow even at 50-digit precision) and are excluded from the analysis.
For \texttt{ext\_log\_maths}, this numerical fingerprinting reduces the $k \leq 8$ $x$-dependent catalogue from 319{,}128 functions to 222{,}838 unique hashes, a reduction of $\sim$30\% (for \texttt{trig\_maths}, the analogous reduction is $\sim$4\%). The rate is particularly high for \texttt{ext\_log\_maths} because there are many syntactically distinct but functionally equivalent ways to compose \texttt{log} and \texttt{Abs}.
Note that 
expressions related by reparametrisation (e.g.\ $a_0 x$ vs.\ $a_0 a_1 x$, which are the same parametric family but evaluate differently at fixed parameter values) are not caught by this procedure.
We estimate this affects $\lesssim 1\%$ of functions (Section~\ref{sec:discussion}); the result is that $\rhok$ will in reality be slightly higher than we measure.

Grouping functions by their derivative fingerprint produces \emph{equivalence classes}: each class contains all primitives $F$ sharing a common integrand $f = F'$. These classes range in size from singletons (integrands with a unique primitive in the grammar) to clusters containing over $10^3$ members.

\subsection{Computing the integrability fraction and equivalence classes}
\label{sec:computing_rho}

Given the fingerprints, we compute $\rhok$ as follows. Let $\mathcal{H}_F = \{\mathrm{hash}(G) : G \in \Fleq\}$ be the set of all function fingerprints, and let $\mathrm{hash}(F')$ denote the fingerprint of the derivative of $F$. Then $F$ is integrable within $\Fleq$ if and only if $\mathrm{hash}(F') \in \mathcal{H}_F$, and $\rhok$ is the fraction of functions satisfying this condition.
Since each function either does or does not have its derivative in $\Fleq$, we report binomial counting uncertainties $\sqrt{\rho(1-\rho)/n}$ with $n = |\Fleq|$, noting that these do not include systematic uncertainties from the imperfect deduplication procedure. Differentiating and hashing $\sim 10^5$ functions takes $\sim 10$ minutes on a single core, while the hash-lookup step (computing $\rhok$ and building equivalence classes) is effectively instantaneous.

\subsection{Comparison with computer algebra systems}
\label{sec:cas_method}

To identify functions where ESI uniquely identifies the antiderivative, we test all multi-member equivalence classes against four CAS chosen to span three fundamentally different algorithmic paradigms:
\begin{enumerate}
\item \emph{Partial algorithmic (Risch-based):} SymPy \citep{Meurer2017} implements a partial Risch algorithm \citep{Risch1969} with heuristic fallbacks. It handles simple towers of exponentials and logarithms but returns unevaluated for deeper compositions.
\item \emph{Heuristic/mixed:} Mathematica's built-in \texttt{Integrate} \citep{Mathematica} combines pattern matching, lookup tables, and algorithmic methods including a partial Risch implementation.
In practice, most heuristic-mixed engines (Mathematica, Maple, SymPy's \texttt{heurisch}) call a variant of the Risch--Norman algorithm \citep{Moses1971,NormanMoore1977} first --- an ansatz-based parallel simplification of Risch that posits a closed-form template, differentiates it, and solves the resulting linear system for the undetermined coefficients. This is fast but incomplete; its failures trigger fallback strategies or return the integral unevaluated.
\item \emph{Rule-based:} RUBI \citep{RichScheibeAbbasi2018,Rubi2024} (Rule-Based Integration) relies exclusively on a curated decision tree of approximately 6{,}700 pattern-matching rules, implemented as a Mathematica package.
\item \emph{Full algorithmic (Risch-based):} FriCAS \citep{FriCAS2024}, a fork of the Axiom computer algebra system, contains the most complete open-source implementation of the Risch algorithm for towers of exponential and logarithmic extensions prevalent in the exhaustive function sets. It serves as the strongest available test of whether an integral is beyond the reach of the Risch algorithm as currently implemented, as opposed to merely beyond heuristic or rule-based methods.
\end{enumerate}
For integrals that resist all four, we additionally test Maxima \citep{Maxima2024} and Giac \citep{Giac2024} via SageMath. An integral that resists all six has therefore been tested against partial and near-complete Risch implementations, commercial heuristics, an independent rule-based system, and two further general-purpose CAS.
This diversity matters because practical symbolic integration is not a single algorithmic path: recent work on Maple treats integration as an algorithm-selection problem and uses machine learning to predict which internal integration routine is likely to succeed \citep{BarketEG2024Selection}. Our use of multiple engines and strategies is a more manual way of probing the same issue. While not exhaustive, our strategy gives us confidence that functions we find to be CAS-resistant cannot be integrated by any current implementation.

We begin by stripping the \texttt{Abs}/\texttt{sign} wrappers that ESR inserts around arguments of \texttt{log}, \texttt{sqrt}, and \texttt{pow} as a precaution during tree enumeration. For SymPy we then apply simplification, try integration with \texttt{conds=`none'} and positive parameter assumptions, and fall back to SymPy's \texttt{manual=True} step-by-step mode if necessary. For Mathematica we pass \texttt{Assumptions -> \{x > 0, params > 0\}} in the initial sweep to match the positive evaluation domain used for fingerprinting and to choose a single real branch for logarithms and radicals; we then re-test the main failures without this assumption and under alternative domains. For RUBI we call \texttt{Int[expr, x]}. For FriCAS, we convert the integrand to FriCAS's InputForm syntax (converting \texttt{atan2} to FriCAS's corresponding \texttt{atan} form) and call \texttt{integrate(expr, x)}. As a cross-check, we also differentiate the known ESI primitive within FriCAS and attempt to integrate the result to avoid any artifacts of the SymPy-to-FriCAS syntax conversion.
Each integral is given a 180-second timeout per strategy in the initial sweep, extended to 600 seconds with additional strategy combinations in the stress test (Section~\ref{sec:cas}). Table~\ref{tab:sizes} reports both stages: ``hard'' counts integrands not solved in the initial sweep, while ``imposs.'' counts those unsolved after the longer timeout and a quasi-exhaustive set of strategies. For computational tractability we test only multi-member equivalence classes.

\section{Results}
\label{sec:results}

\subsection{The integrability fraction $\rhok$}
\label{sec:rho}

The central quantity in this study is the \emph{integrability fraction} (Eq.~\ref{eq:1}), which in the notation of Section~\ref{sec:method} can be written as
\begin{equation}
\label{eq:rho}
\rhok \;=\; \frac{\bigl|\bigl\{F \in \Fleq : \mathrm{hash}(F') \in \mathrm{hash}(\Fleq)\bigr\}\bigr|}{|\Fleq|}\,.
\end{equation}
This is the proportion of functions in the enumerated space whose derivatives also belong to that space.
A function $F$ contributes to the numerator if and only if there exists some $G \in \Fleq$ whose numerical fingerprint matches that of $F'$; equivalently, $F$ is integrable within the grammar if there is a closed-form antiderivative expressible at complexity $\leq k$ in the same operator basis.

\begin{figure}[t]
\centering
\includegraphics[width=\columnwidth]{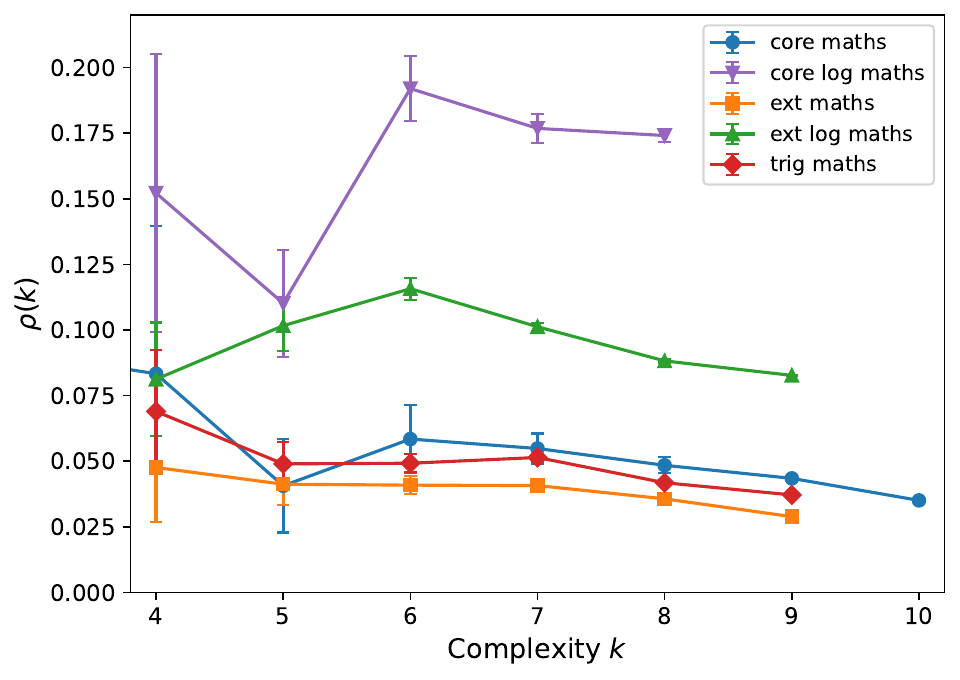}
\caption{Integrability fraction $\rhok$ as a function of complexity $k$ for all five operator bases. Error bars are binomial ($\sqrt{\rho(1-\rho)/n}$). The two log-containing bases (\texttt{core\_log\_maths} and \texttt{ext\_log\_maths}) exhibit elevated $\rhok$ and a pronounced non-monotonic peak at $k=6$.}
\label{fig:rho}
\end{figure}

Fig.~\ref{fig:rho} shows $\rhok$ for each basis, revealing two key patterns. First, the bases with high-complexity coverage exhibit a declining $\rhok$: \texttt{core\_maths} falls from $\sim 5.5\%$ at $k = 7$ to $3.5\%$ at $k = 10$, \texttt{ext\_maths} from $\sim 4\%$ at $k = 5$--$7$ to $2.9\%$ at $k=9$ and \texttt{trig\_maths} from $\sim 5\%$ to $3.7\%$ over the same range. Second, the behaviour of \texttt{ext\_log\_maths} is strikingly different from the non-log bases: it exhibits a clear non-monotonic pattern, rising from 8.1\% at $k=4$ to a peak of 11.6\% at $k=6$ before declining monotonically to 8.3\% at $k=9$. The \texttt{core\_log\_maths} basis, run to $k=8$, shows the same elevated log-driven integrability.

Adding the logarithm boosts $\rhok$ by a factor of $\sim 2.5$ for $k\geq 5$ in the \texttt{ext\_log\_maths}/\texttt{ext\_maths} comparison.
The effect of the logarithm can be isolated from possible interactions with other operators by examining the \texttt{core\_log\_maths} basis containing only \texttt{inv} and \texttt{log} as unary operators (i.e.\ \texttt{core\_maths} plus \texttt{log}, without \texttt{exp}, \texttt{sqrt} or \texttt{sq}). This basis also exhibits a peak at $k=6$ ($\rho = 19.2\%$) and elevated integrability at every complexity ($3.6\times$ above \texttt{core\_maths} at $k=8$), confirming that the logarithm alone drives the enhancement---no interaction with the exponential is required. The higher absolute values ($\rho = 17.4\%$ at $k=8$ vs $8.8\%$ for \texttt{ext\_log\_maths}) reflect the smaller operator set: adding \texttt{exp}, \texttt{sqrt} and \texttt{sq} introduces many non-integrable functions that dilute the fraction.
The mechanism behind the logarithm's enhancement is analysed in Section~\ref{sec:mechanism}.

Perhaps surprisingly, the trig\_maths basis does not exhibit a similarly elevated $\rho(k)$ despite the fact that $\sin$ and $\cos$ form a closed orbit under differentiation ($\dd/\dd x[\sin f] = f'\cos f$, $\dd/\dd x[\cos f] = -f'\sin f$): the $\rhok$ values for trig\_maths are comparable to those of ext\_maths, not ext\_log\_maths. This is because the chain rule introduces a multiplicative factor of $f'$ that drives rapid complexity growth, offsetting the closure of the trigonometric pair.
That this has a similar $\sim 4$--$5\%$ integrability fraction while not being nested in any other basis set indicates that this is not an artifact of the nesting \texttt{core\_maths} $\subset$ \texttt{ext\_maths} $\subset$ \texttt{ext\_log\_maths}.
It is important to note however that \texttt{trig\_maths}'s structural independence from the other bases is at the level of the operator grammar not the function field itself, since $\sin$ and $\cos$ can be expressed via $\exp$ in the complex plane.

\subsection{Why log matters}
\label{sec:mechanism}

\begin{figure}[t]
\centering
\includegraphics[width=\textwidth]{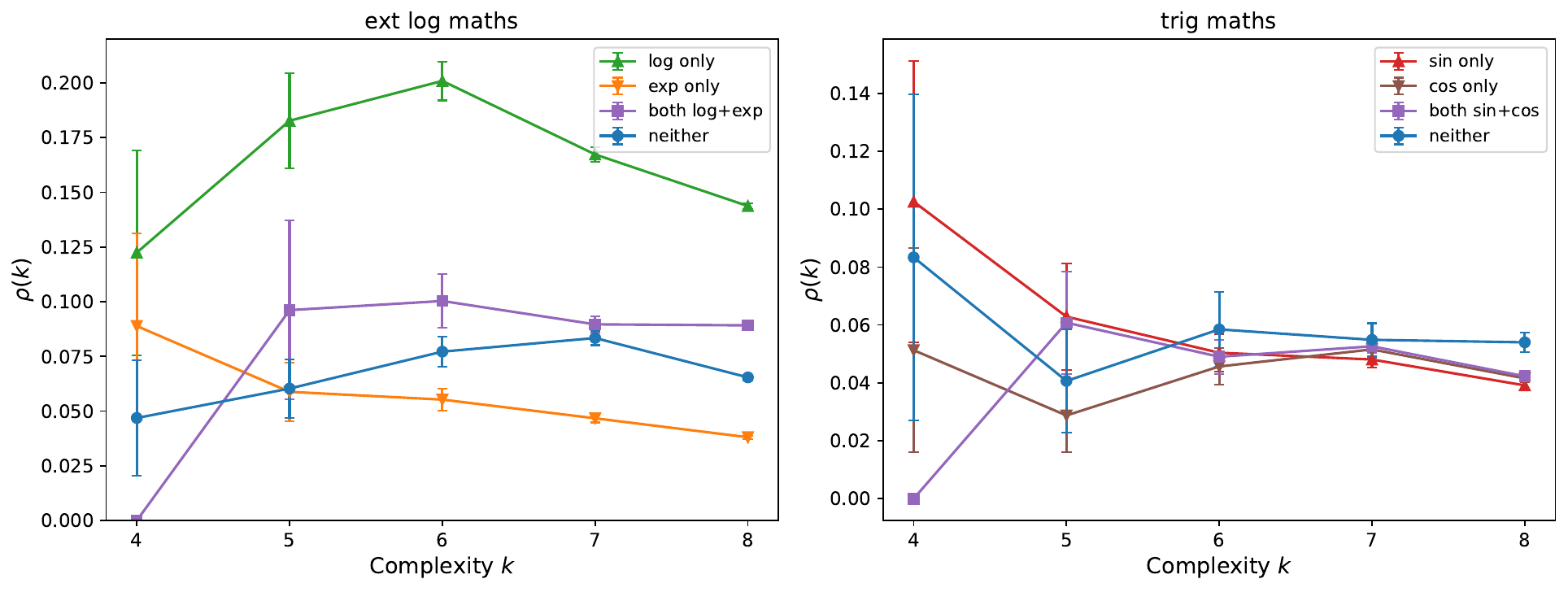}
\caption{Decomposition of $\rhok$ by operator presence, with binomial error bars.  \emph{Left:} \texttt{ext\_log\_maths}, partitioned by log/exp content.  The log-only subset dominates, with $\rho$ peaking at $20.1\%$ at $k=6$.  \emph{Right:} \texttt{trig\_maths}, partitioned by sin/cos content, where no comparable asymmetry is observed.}
\label{fig:decomp}
\end{figure}

Figure~\ref{fig:decomp} decomposes $\rhok$ by partitioning $\Fleq$ into subsets according to which transcendental operators each function contains. The left panel shows the \texttt{ext\_log\_maths} basis: functions containing \texttt{log} but not \texttt{exp} (``log only'') are by far the most integrable, with $\rho$ peaking at $20.1\%$ at $k=6$; functions containing only \texttt{exp} are the \emph{least} integrable. The right panel shows the \texttt{trig\_maths} basis, where no comparable asymmetry between \texttt{sin} and \texttt{cos} is observed.

The subset containing logarithms but not exponentials 
is
the most integrable at every complexity level: at $k = 8$ its integrability
fraction of $14.4\%$ exceeds that of the exp\_only subset by a factor of $3.8$.
Functions containing any logarithm are $2.7$ times more likely to be
integrable than those without.  Most strikingly, the exponential alone
\emph{diminishes} integrability: the ``exp only'' subset has the lowest $\rho$ of
any category, falling below even the ``neither'' subset that contains no
transcendental operators at all.  This rules out any explanation based on
compositional enrichment---the logarithm's effect is not merely to enlarge the
function space, but to preferentially generate derivative forms that close back onto functions already present.
The complexity gap $\delta \equiv \min\text{-complexity}(F') - \text{complexity}(F)$ is useful as a diagnostic but is not by itself the explanation. For \texttt{ext\_log\_maths} at $k\leq8$ the mean $\delta = -1.13$ and $89.5\%$ of functions satisfy $\delta\leq0$, so differentiation typically \emph{reduces} expression complexity. Log-containing functions shift this only slightly (mean $\delta=-1.20$, $89.9\%$ with $\delta\leq0$ vs $-1.03$ and $88.6\%$ for functions without log); the stronger evidence for the logarithm's special role is the operator decomposition in Figure~\ref{fig:decomp}.

The peak in the \texttt{ext\_log\_maths} $\rhok$ arises because the space of distinct derivative forms grows faster than the function space. As $k$ increases, derivatives become increasingly varied and structurally specific, so the probability of matching an existing function declines. The logarithm delays this onset because the chain rule $\dd/\dd x[\log(f)] = f'/f$ produces rational forms that overlap efficiently with the existing function catalogue: at $k=6$, the measured integrability fraction is 20.1\% for the \texttt{log\_only} subset and 5.5\% for the \texttt{exp\_only} subset.
The magnitude of this asymmetry can be understood as follows. The derivative $f'/f$ is a rational function whose ``target space'' is the entire set of rational expressions in~$\Fleq$---a large subset. By contrast, $f'\exp(f)$ necessarily contains an exponential factor and can therefore only match functions in $\Fleq$ that also contain~$\exp$, a much smaller target. Moreover, the product $f'\exp(f)$ has complexity at least $|f| + |f'| + 1$, which for all but the simplest $f$ exceeds the complexity ceiling, further reducing the matching probability. The factor of $3.6$ thus reflects the ratio of effective target-space sizes rather than a subtle cancellation, though a quantitative derivation from the grammar remains an open problem (\ref{app:model}). The empirical growth rates of the function and derivative-form spaces, together with a partial analytical derivation from the operator grammar, are presented in \ref{app:model}.

The derivative map also has a strongly clustered structure. Grouping functions by the numerical fingerprint of $F'$ produces equivalence classes of primitives with a common integrand. For \texttt{ext\_log\_maths} at $k \leq 8$, 49{,}538 of the 203{,}812 clusters are multi-member, while \texttt{trig\_maths} has only 6{,}987 multi-member clusters out of 131{,}734. The largest clusters correspond to attractor integrands such as $1$, $-1$ and $1/x$, which accumulate many syntactically distinct primitives.

\subsection{Comparison with computer algebra systems}
\label{sec:cas}

As a preliminary validation, we verify that ESI correctly handles known non-elementary integrands. Five classical non-elementary integrands \citep[see e.g.][]{Bronstein2005} --- $e^{x^2}$, $e^{e^x}$, $e^x / x$, $x^x$, and $e^{\sqrt{x}}$ --- are correctly absent from the derivative database, while their integrable variants (e.g.\ $2x\,e^{x^2}$, the derivative of $e^{x^2}$) are correctly found with dozens of primitive representations, consistent with Risch--Liouville theory \citep{Risch1969,Bronstein2005}.

We test all 49{,}538 multi-member equivalence classes in \texttt{ext\_log\_maths} against the four CAS engines described in Section~\ref{sec:cas_method}. In the initial 180-second sweep, SymPy successfully integrates 47{,}107 of the 49{,}538 \texttt{ext\_log\_maths} integrands (95.1\%). The remaining 2{,}431 (4.9\%) are genuine failures: not timeouts, but cases where SymPy's integration algorithms explicitly return unevaluated integrals, typically within seconds.
With cleaned input and positive-domain assumptions, Mathematica successfully integrates 47{,}489 of 49{,}538 integrands (95.9\%), with 2{,}049 failures (4.1\%). Cross-matching the two failure sets reveals 503 integrals that \emph{both} SymPy and Mathematica fail on at 180 seconds.

To distinguish ``hard'' integrals (solvable with more time or alternative strategies) from genuinely impossible ones, we performed a comprehensive stress test on all 503 \texttt{ext\_log\_maths} integrals failed by both SymPy and Mathematica at 180 seconds. Each was tested with nine SymPy strategy combinations and three Mathematica strategies, each with a 600-second timeout per strategy. Of these 503, 270 (53.8\%) are solved by at least one strategy, while 232 resist both SymPy and Mathematica even at 600\,s. Those 232 were then passed to FriCAS, which solves 218.
This shows that most failures of SymPy, Mathematica and RUBI are implementation-coverage gaps rather than evidence that no elementary antiderivative exists. FriCAS solves all $x^x$-type and $f(x)^{g(x)}$ integrals that defeat the other three engines. By contrast, RUBI solves none of the 232 integrals that SymPy and Mathematica leave unsolved; as a pattern library designed for textbook integral forms, it does not aim to cover these deeply nested exp-log towers.
Table~\ref{tab:sizes} summarises the broader per-basis initial and stress-test counts where available.
 
Across the remaining CAS-tested bases, \texttt{core\_maths} produces no all-engine failures; \texttt{ext\_maths} leaves two Mathematica failures after the corrected no-\texttt{Assumptions} rerun; and the 685 \texttt{trig\_maths} Mathematica failures are all solved by Sage (502) or FriCAS (183). A further independent broad search of 133{,}576 Abs-free $k=9$ candidates in \texttt{ext\_log\_maths} via a Mathematica--Sage--FriCAS cascade also leaves zero new survivors, validating the uniqueness of the Abs-free headline. In total, three integrals across the four CAS-tested bases resist all six engines under all tested strategies (Table~\ref{tab:cas_blind}). ESI also identifies two \texttt{ext\_log\_maths} Abs-containing integrands at $k=8$, $\sqrt{\lvert x \pm \lvert\!\log x\rvert^{\log x}\rvert}$, that fail all standard CAS calls but are solved by Mathematica after restricting to $x>1$.
The incomplete entries in Table~\ref{tab:sizes} therefore do not weaken this final count: filling them would require full-basis stress tests for integrals already eliminated from the six-engine survivor pipeline, or, for \texttt{core\_log\_maths}, a basis added to diagnose the log enhancement rather than to extend the CAS-resistant search.

As a separate exercise, we compared ESI against RUBI's own benchmark suite of 72{,}401 test integrals \citep{Rubi2024}. These are curated integrals drawn from textbooks and reference works --- a fundamentally different problem set from the ESI-generated functions discussed above, with zero overlap with the CAS-resistant integrals (which are novel functions from ESI's enumeration that do not appear in any existing test suite or integral table). By mapping the Mathematica-syntax integrands to ESI's numerical fingerprinting scheme, we find that ESI (through the \texttt{ext\_log\_maths} and \texttt{trig\_maths} bases at $k \leq 8$) can solve 184 RUBI benchmark problems. That this fraction is small is expected given ESI's complexity ceiling; however, among these are 7 integrals classified as ``hard'' by RUBI (requiring $\geq 5$ rule applications), including one requiring 13 steps: $\int x \sec x\,(2 + x \tan x)\,\dd x$, which ESI readily identifies as $x^2/\cos x$ at $k = 6$. We also checked RUBI's own failures: of the 134 integrals in RUBI's test suite that RUBI itself cannot evaluate, 122 can be converted from Mathematica syntax to a numerical fingerprint (12 fail the syntax conversion). Of these 122, only 3 have their integrand fingerprint present in ESI's derivative database; the remainder involve functions that exceed ESI's complexity ceiling or use operators outside ESI's bases. One of the 3 matches is a hyperbolic branch artefact, leaving two robust ESI solves:
\begin{equation}
	\int x^{-2 - 1/x}(1 - \log x)\,\dd x = -x^{-1/x} \nonumber
\end{equation}
and
\begin{equation}
	\int \bigl((x+1)\log^2 x - 1\bigr)\,e^{x + 1/\log x}/\log^2 x\,\dd x = x\,e^{x+1/\log x}. \nonumber
\end{equation}
Both of these involve deeply nested log/exp compositions of the kind that fall through the gaps of rule-based systems but lie squarely within ESI's catalogued space. FriCAS also solves them.

\begin{table*}
  \centering
  \caption{The integrals that are solved by ESI but resist all six CAS engines under all tested strategies. All three pair a square-root algebraic extension with a transcendental tower ($e^x$, $e^{-x}$, or $x^x$) whose coupling cannot be undone by a closed-form substitution.}
  \label{tab:cas_blind}
  \small
  \begin{tabular}{clcc}
    \toprule
    $k$ & Basis & Integrand & Antiderivative \\
    \midrule
    9 & \texttt{ext\_log} &
      $\dfrac{(2x+1)\,e^x + 1}{2\sqrt{x + e^{-x}}}$ &
      $e^x\sqrt{x + e^{-x}}$ \\[8pt]
    9 & \texttt{ext} &
      $\dfrac{(2x+1)\,e^x - 1}{2\sqrt{\lvert x - e^{-x}\rvert}}$ &
      $e^x\sqrt{\lvert x - e^{-x}\rvert}$ \\[8pt]
    8 & \texttt{ext} &
      $\dfrac{x^{x - 3/2}\,(1 - x - x\log x)}{2\sqrt{\lvert x - x^x\rvert}}$ &
      $\sqrt{\lvert x - x^x\rvert}\,/\,\sqrt{x}$ \\[8pt]
    \bottomrule
  \end{tabular}
\end{table*}

Besides being able to evaluate integrals, another consideration is the form in which the integrals are provided.
ESI stores the lowest-complexity primitive among the expression trees that ESR explicitly enumerates and recognises as equivalent. CAS returns independently simplified expressions, and can sometimes apply cancellations or branch-specific identities that ESR's symbolic and numerical deduplication did not identify as the same function family. Conversely, ESI often stores compact factored or nested representatives that CAS expands. It is therefore fair to assess which method produces simpler forms on average, while recognising that the comparison is metric-dependent.
Among the 47{,}107 integrals that both ESI and SymPy successfully evaluate, SymPy returns a lower node count in 38.7\% of cases, ESI in 34.7\%, and they are tied in 26.6\%. However, across all integrals where ESI provides the simpler form, ESI saves a total of 135{,}036 nodes, compared to 32{,}431 saved by SymPy when it wins --- a ratio of 4.2:1. ESI's wins are less frequent but substantially larger, reflecting its preference for factored, nested representations that CAS typically expands.

This reveals a complementarity between ESI and CAS, with the value of ESI lying
in its exhaustive structural coverage: it catalogues every integrable function within a defined space, including classes that fall through the gaps of general-purpose algorithms.
On GitHub we provide a Python script and Jupyter notebook for solving integrals by ESI through a lookup to the precomputed derivative database.

\section{Discussion}
\label{sec:discussion}

\subsection{Integrability fraction}

Liouville's theorem \citep{Liouville1835,Rosenlicht1972}
characterises the constrained form of elementary antiderivatives when they exist, but does not say anything about the complexity of these integrals. The integrability fraction $\rhok$ that we introduce fills in this information, measuring the proportion of functions at complexity $k$ whose antiderivatives are also expressible at complexity $\leq k$ within the same operator basis. The declining $\rhok$ observed in the bases with high-complexity coverage (Section~\ref{sec:rho}) shows that the
antiderivatives increasingly require higher complexity to express and hence increasingly fall outside the enumerated space. The $\sim 2.5\times$ elevation of $\rhok$ in \texttt{ext\_log\_maths} relative to \texttt{ext\_maths} for $k\geq 5$ reflects the approximate closure of the exp--log differential subsystem: the logarithm's derivative rule $\dd/\dd x[\log f] = f'/f$ produces rational forms that frequently match existing functions within the enumerated space, whereas the exponential's derivative $f'\exp(f)$ tends to increase complexity beyond the enumeration boundary. A quantitative model formalising this asymmetry is presented in~\ref{app:model}.

More broadly, the $\rhok$ framework suggests a Galois-theoretic formulation. The Kolchin--Picard--Vessiot theory of differential field extensions \citep{Magid1994} provides a natural algebraic setting in which to define ``closure densities'' on isomorphism classes of extensions of given transcendence degree, which would quotient out the dependence on the specific operator grammar and complexity metric. Whether the $\sim$2--3 enhancement of $\rhok$ by the logarithm corresponds to a structural feature of the exp--log differential subfield, or is dependent on our counting apparatus, is a question that such a reformulation could in principle answer.

\subsection{Novel antiderivatives and comparison with CAS}

The Risch algorithm \citep{Risch1969,Bronstein2005} is fully decision-procedural for elementary functions in principle, but current implementations remain incomplete for algebraic extensions over exponential--logarithmic towers \citep{Trager1976,Bronstein2005}. FriCAS \citep{FriCAS2024} comes closest; its remaining failures point to specific incomplete subroutines such as constant residues and polynomial parts.
The CAS comparison of Section~\ref{sec:cas} can therefore be seen as diagnosing incompleteness in current Risch implementations.

The three integrals in Table~\ref{tab:cas_blind} share a structural pattern: a square-root algebraic extension coupled to $e^x$, $e^{-x}$, or $x^x$ in a way that cannot be undone by a closed-form substitution. The Abs-free headline $e^x\sqrt{x+e^{-x}}$ is the clearest example; the natural substitution $u=x+e^{-x}$ leaves the external $e^x$ inaccessible without inverting a transcendental equation. The two \texttt{ext\_maths} integrals share the same structural signature. ESI also identifies two further \texttt{ext\_log\_maths} Abs-containing integrands ($\sqrt{\lvert x \pm \lvert\!\log x\rvert^{\log x}\rvert}$, $k=8$) that fail all standard CAS calls but are solved by Mathematica after restricting to $x>1$.

Integration involving fractional powers of logarithms also connects to the theory of integration in terms of special functions: \citet{Cherry1985,Cherry1986} developed algorithms for integration via $\operatorname{Ei}$ and $\operatorname{erf}$, and \citet{Hebisch2018} extended these to incomplete gamma functions $\Gamma(a, x)$ with rational $a$; more broadly, \citet{Raab2012,Raab2013} generalise the Risch framework to differential fields containing non-elementary monomials (\textrm{erf}, incomplete gamma, polylogarithms), and \citet{KauersKoutschan2015,Koutschan2013} develop holonomic
integration, which targets antiderivatives that satisfy linear ODEs with polynomial coefficients rather than admitting closed elementary forms. Our class~(i) integrals ($e^{cx} (\log x)^{1/2}$ and similar) \emph{might} in principle be expressible via $\Gamma(1/2, \ldots)$, but the antiderivatives ESI finds are elementary --- the CAS should find them without recourse to special functions but cannot because the algebraic extension over the tower triggers unimplemented code paths. To our knowledge, no prior work has systematically catalogued elementary integrals that expose these implementation gaps; the closest is the independent CAS benchmarking of \citet{Nasser2024}, which tests existing collections of integrals across multiple engines but does not generate new ones or target specific algorithmic weaknesses.

To check for prior appearances of the Abs-free headline, we searched for $e^x\sqrt{x + e^{-x}}$ and its derivative in standard integral tables \citep{GradshteynRyzhik2015,Prudnikov1986,DLMF}, in the 106{,}812-integral independent CAS benchmark suite of \citet{Nasser2024}, and in mathematical databases and forums; we found no occurrence.
We therefore present, to our knowledge, the first algorithm that finds the antiderivatives in Table~\ref{tab:cas_blind} automatically. Note that this does not mean that finding the antiderivatives is fundamentally hard: they can be guessed with a little trial and error, and modern large language models can also identify them.

\subsection{Limitations and further work}

The main limitation of our $\rho(k)$ results is that they depend on ESR's specific expression grammar, which determines how the function space is partitioned by complexity. Changing the grammar or the complexity metric changes both the denominator of $\rhok$ and the location of features such as the \texttt{ext\_log\_maths} peak. Indeed, the non-monotonic peak at $k=6$ in the ESR node-count metric does not appear under alternative complexity functionals such as expression-tree depth, operator count, or reassignment of each function to the minimum complexity at which its numerical fingerprint first appears. For example, if the node-count metric is modified so that each \texttt{log} or \texttt{exp} operator contributes 2 rather than 1 unit of complexity, $\rhok$ instead flattens to $\sim 8$--$9\%$ with no discernible peak.

The quantitative results also depend on the deduplication method: under string-only deduplication, the peak disappears because syntactic identity misses the additional equivalences found by numerical fingerprinting. Future grammar-level improvements, including hash-based simplification methods \citep{hash,Kronberger2024}, may therefore shift the quantitative values and should be treated as a change in the measured function space.
A further subtlety is that differentiation can produce compound parameter expressions --- for example, $F = a_0 e^{a_1 x}$ has $F' = a_0 a_1 e^{a_1 x}$, which is structurally the same parametric family as $a_0 e^{a_1 x}$ but carries an extra factor of $a_1$ that prevents the numerical fingerprint from matching. This causes $\rhok$ to slightly undercount integrable functions and hence our integrability fraction to be a lower bound, although we estimate that $\lesssim 1\%$ of functions are affected based on counting derivatives containing parameter products $a_i a_j$.
The stable result is not the precise peak location but the broader log enhancement and high-complexity decline within the tested grammar.

There are several natural extensions to this analysis. First, the complexity ceiling of $k \leq 8$--$10$ restricts the analysis to relatively simple expressions; many primitives of practical interest lie at higher complexity. An upcoming upgrade (Kronberger et al 2026, in prep) will extend ESR to a few complexity units higher for the same computational cost. Second, a combined basis incorporating trigonometric, exponential, and logarithmic operators would probe the interplay between all three approximately closed differential subsystems. Introducing special functions such as $\operatorname{erf}$ or $\operatorname{Ei}$ would extend the analysis beyond the elementary functions. Third, an analytical derivation of $\alpha$ and $\beta$ from the operator grammar---potentially via analytic combinatorics \citep{FlajoletSedgewick2009} of labelled trees---would elevate the growth-rate crossover from an empirical observation to a predictive framework.
Finally, a systematic study of the $f^g$ integral class --- characterising which pairs $(f, g)$ produce derivatives that CAS can and cannot evaluate --- could reveal the algebraic obstructions that prevent current implementations of the Risch algorithm from handling towers of exponentials and logarithms, which could in turn inform the construction of superior algorithms in the future.

\section{Conclusion}
\label{sec:conclusion}

We present a concrete implementation of integration by differentiation: the Exhaustive Symbolic Integration (ESI) algorithm. ESI builds exhaustive function sets composed of operators in a user-defined basis set up to a maximum complexity $k$, then differentiates those functions to define a function--antiderivative mapping. This affords investigation of the landscape of symbolic integration---the frequency with which integration preserves the operator set and complexity limit---and the identification of antiderivatives that cannot be found by computer algebra systems.

Our central findings are twofold. First, the operator basis profoundly shapes the integrability landscape: adding the logarithm boosts within-class integrability fraction $\rhok$ by factors of order $2$--$3$ over most shared complexity levels. Bases with high-complexity coverage show a declining $\rhok$ at $k = 9$--$10$, with \texttt{ext\_maths} dropping to $2.9\%$ and \texttt{core\_maths} to $3.5\%$, because the space of distinct derivative forms grows faster than the function space (\ref{app:model}). This is part of the broader quantification of the integrability landscape across the five operator bases studied here, although we caution that the quantitative results depend on the expression grammar and complexity metric.

Second, ESI systematically discovers integrals that expose gaps in existing CAS implementations.
In particular we identify
three integrals in the four bases included in the full CAS cascade that resist all tested CAS engines (SymPy, Mathematica, RUBI, FriCAS, Maxima, and Giac) under all tested strategies. These form a structurally coherent family pairing an algebraic extension $\sqrt{\cdot}$ with a transcendental tower ($e^x$, $e^{-x}$, or $x^x$) whose coupling cannot be inverted by a closed-form substitution. We also solve two integrals from the RUBI benchmark set that RUBI itself cannot evaluate.

ESI's value lies not in competing with CAS on integration in general, but in its exhaustive coverage: it catalogues every integrable function within the grammar, enabling for the first time both $\rhok$ measurement and the systematic discovery of integrals beyond the capabilities of existing algorithms. Extending the method to higher complexity and alternative basis sets would test the robustness of the integrability patterns found here, discover further antiderivatives, and broaden the systematic exploration of limitations in existing Risch implementations with an eye to shoring them up in the future.

\section*{Data availability}

The code underlying this article is available at \url{https://github.com/harrydesmond/ExhaustiveSymbolicIntegration}; this repository contains the ESI pipeline, CAS-comparison scripts, figure-generation scripts, paper source, and the \texttt{esi\_integrate.py} lookup tool for finding integrals through ESI, with a worked notebook demonstration. The accompanying Zenodo record \url{https://doi.org/10.5281/zenodo.20027938} is the data deposit, containing the products used for the paper tables and figures, lookup-equivalence databases and raw derivative/fingerprint outputs, and the catalogues needed to regenerate the pipeline inputs.

\section*{Acknowledgements}

I thank Nicolas Tessore for the discussion that sparked this project. I am supported by a Royal Society University Research Fellowship (grant no. 211046).

\appendix

\section{Growth-rate crossover model}
\label{app:model}

This appendix examines how fast the function space and derivative-form space grow with complexity, and what this implies for the decline of $\rhok$. The key question is whether the growth rates can be derived from the operator grammar or must be measured empirically. We show that the raw tree count (before deduplication) follows from standard analytic combinatorics, but that the effective growth rate after deduplication and the growth rate of the derivative space cannot; we therefore present their measurements from empirical fits.

We first note that the \emph{raw} tree count --- before any deduplication --- can be derived exactly from the operator grammar via analytic combinatorics \citep{FlajoletSedgewick2009}. Each basis defines a context-free tree grammar with $p_0$ leaf types (always 2: $x$ and $a$), $p_1$ unary operators, and $p_2$ binary operators (always 5: $+, -, \times, \div, \mathrm{pow}$). The generating function for the number of trees of size $k$ satisfies
\[
T(z) = p_0\,z + p_1\,z\,T(z) + p_2\,z\,T(z)^2\,,
\]
whose dominant singularity $r$ gives the raw growth rate $\alpha_{\mathrm{raw}} = 1/r$ via the Drmota--Lalley--Woods theorem \citep{FlajoletSedgewick2009}. Solving the discriminant condition yields
\[
\alpha_{\mathrm{raw}} = p_1 + 2\sqrt{p_0\,p_2} = p_1 + 2\sqrt{10}\,,
\]
with an asymptotic tree count $T_k \sim C\,\alpha_{\mathrm{raw}}^k\,k^{-3/2}$, where the $k^{-3/2}$ subexponential factor is universal for this grammar class. The resulting values are:

\begin{center}
\begin{tabular}{lcccc}
\toprule
& \texttt{core} & \texttt{ext} & \texttt{ext\_log} & \texttt{trig} \\
\midrule
$p_1$ & 1 & 4 & 5 & 3 \\
$\alpha_{\mathrm{raw}}$ & 7.3 & 10.3 & 11.3 & 9.3 \\
$\alpha_{\mathrm{eff}}$ (measured) & 3.2 & 5.7 & 6.1 & 5.5 \\
\bottomrule
\end{tabular}
\end{center}

The ordering $\alpha_{\mathrm{raw}}(\mathrm{core}) < \alpha_{\mathrm{raw}}(\mathrm{trig}) < \alpha_{\mathrm{raw}}(\mathrm{ext}) < \alpha_{\mathrm{raw}}(\mathrm{ext\_log})$ matches the observed ordering of empirical growth rates, confirming that larger operator sets produce faster-growing function spaces. However, $\alpha_{\mathrm{eff}}$ is roughly half of $\alpha_{\mathrm{raw}}$ in every case, because ESR's symbolic deduplication and our numerical fingerprinting remove 75--98\% of raw trees as duplicates (Section~\ref{sec:fingerprint}). The effective growth rate depends on the algebraic identities among the specific operators in the basis (e.g.\ $\log(\exp(x)) = x$, commutativity of addition), which cannot be captured by a grammar-level analysis alone.

We next describe the empirical growth-rate patterns. The function space grows as $|\mathcal{F}_k| \sim \alpha^k$ and the space of distinct derivative forms as $|S_k| \sim \beta^k$. The measured growth rates are $\alpha = 3.2$, $5.7$, $6.1$, $5.5$ and $\beta = 4.0$, $6.0$, $7.0$, $7.0$ for \texttt{core\_maths}, \texttt{ext\_maths}, \texttt{ext\_log\_maths} and \texttt{trig\_maths} respectively. The \texttt{core\_log\_maths} basis is omitted from these regression fits because its high-complexity derivative-form sequence has not been run to the same depth. In every case $\beta > \alpha$, which is expected: the chain and product rules generically increase structural diversity, so derivative forms proliferate faster than the functions that generate them. This guarantees the eventual decline of $\rhok$ at high complexity, regardless of the complexity metric used. Deriving $\beta$ analytically is substantially harder than $\alpha_{\mathrm{raw}}$, because differentiation is not a grammar-preserving operation: the chain rule multiplies subtrees and the product rule duplicates structure, so the derivative-form space is not generated by a context-free grammar. We therefore report $\beta$ as an empirical fit.

The ratio $\alpha/\beta < 1$ sets the asymptotic decline rate of $\rhok$. The measured values are $\alpha/\beta = 0.81$, $0.94$, $0.87$, $0.78$ for the four fitted bases respectively. We caution that these ratios should be interpreted as empirical regression fits describing the high-$k$ decline, not as a mechanistic model of $\rhok$. In particular, a monotonic ratio $(\alpha/\beta)^k$ cannot produce the non-monotonic peak observed in \texttt{ext\_log\_maths}; the peak likely reflects transient effects such as ``attractor'' integrands (low-complexity derivatives like $1$, $1/x$, $-1$ that accumulate many primitives at moderate $k$) and basis-dependent deduplication rates that vary non-monotonically with $k$. A full analytical treatment would require extending the generating-function approach to tree transducers representing the differentiation operator \citep{FlajoletSedgewick2009}.

\bibliographystyle{elsarticle-harv}
\bibliography{references}

\begin{thebibliography}{39}
\expandafter\ifx\csname natexlab\endcsname\relax\def\natexlab#1{#1}\fi
\providecommand{\url}[1]{\texttt{#1}}
\providecommand{\href}[2]{#2}
\providecommand{\path}[1]{#1}
\providecommand{\DOIprefix}{doi:}
\providecommand{\ArXivprefix}{arXiv:}
\providecommand{\URLprefix}{URL: }
\providecommand{\Pubmedprefix}{pmid:}
\providecommand{\doi}[1]{\href{http://dx.doi.org/#1}{\path{#1}}}
\providecommand{\Pubmed}[1]{\href{pmid:#1}{\path{#1}}}
\providecommand{\bibinfo}[2]{#2}
\ifx\xfnm\relax \def\xfnm[#1]{\unskip,\space#1}\fi
\bibitem[{Abbasi(2024)}]{Nasser2024}
\bibinfo{author}{Abbasi, N.M.}, \bibinfo{year}{2024}.
\newblock \bibinfo{title}{Computer algebra independent integration tests}.
\newblock \bibinfo{note}{Available at
  \url{https://www.12000.org/my_notes/CAS_integration_tests/}. 72{,}000+ test
  integrals compared across CAS}.
\bibitem[{Barket et~al.(2023)Barket, England and Gerhard}]{BarketEG2023}
\bibinfo{author}{Barket, R.}, \bibinfo{author}{England, M.},
  \bibinfo{author}{Gerhard, J.}, \bibinfo{year}{2023}.
\newblock \bibinfo{title}{Generating elementary integrable expressions}, in:
  \bibinfo{booktitle}{Computer Algebra in Scientific Computing (CASC)},
  \bibinfo{publisher}{Springer}. pp. \bibinfo{pages}{21--38}.
\newblock \DOIprefix\doi{10.1007/978-3-031-41724-5_2},
  \href{http://arxiv.org/abs/2306.15572}{{\tt arXiv:2306.15572}}.
\bibitem[{Barket et~al.(2024a)Barket, England and Gerhard}]{BarketEG2024}
\bibinfo{author}{Barket, R.}, \bibinfo{author}{England, M.},
  \bibinfo{author}{Gerhard, J.}, \bibinfo{year}{2024}a.
\newblock \bibinfo{title}{The {Liouville} generator for producing integrable
  expressions}, in: \bibinfo{booktitle}{Computer Algebra in Scientific
  Computing (CASC)}, \bibinfo{publisher}{Springer}. pp.
  \bibinfo{pages}{47--62}.
\newblock \DOIprefix\doi{10.1007/978-3-031-69070-9_4},
  \href{http://arxiv.org/abs/2406.11631}{{\tt arXiv:2406.11631}}.
\bibitem[{Barket et~al.(2024b)Barket, England and
  Gerhard}]{BarketEG2024Selection}
\bibinfo{author}{Barket, R.}, \bibinfo{author}{England, M.},
  \bibinfo{author}{Gerhard, J.}, \bibinfo{year}{2024}b.
\newblock \bibinfo{title}{Symbolic integration algorithm selection with machine
  learning: {LSTM}s vs tree {LSTM}s}, in: \bibinfo{booktitle}{Mathematical
  Software -- {ICMS} 2024}, \bibinfo{publisher}{Springer}. pp.
  \bibinfo{pages}{167--175}.
\newblock \DOIprefix\doi{10.1007/978-3-031-64529-7_18}.
\bibitem[{Bartlett et~al.(2024)Bartlett, Desmond and Ferreira}]{Bartlett2024}
\bibinfo{author}{Bartlett, D.J.}, \bibinfo{author}{Desmond, H.},
  \bibinfo{author}{Ferreira, P.G.}, \bibinfo{year}{2024}.
\newblock \bibinfo{title}{Exhaustive symbolic regression}.
\newblock \bibinfo{journal}{IEEE Transactions on Evolutionary Computation}
  \bibinfo{volume}{28}, \bibinfo{pages}{950--964}.
\newblock \DOIprefix\doi{10.1109/TEVC.2023.3280250},
  \href{http://arxiv.org/abs/2211.11461}{{\tt arXiv:2211.11461}}.
\bibitem[{Bronstein(2005)}]{Bronstein2005}
\bibinfo{author}{Bronstein, M.}, \bibinfo{year}{2005}.
\newblock \bibinfo{title}{Symbolic Integration {I}: Transcendental Functions}.
  volume~\bibinfo{volume}{1} of \textit{\bibinfo{series}{Algorithms and
  Computation in Mathematics}}.
\newblock \bibinfo{edition}{2nd} ed., \bibinfo{publisher}{Springer},
  \bibinfo{address}{Berlin, Heidelberg}.
\newblock \DOIprefix\doi{10.1007/b138171}.
\bibitem[{{Burlacu} et~al.(2021){Burlacu}, {Kammerer}, {Affenzeller} and
  {Kronberger}}]{hash}
\bibinfo{author}{{Burlacu}, B.}, \bibinfo{author}{{Kammerer}, L.},
  \bibinfo{author}{{Affenzeller}, M.}, \bibinfo{author}{{Kronberger}, G.},
  \bibinfo{year}{2021}.
\newblock \bibinfo{title}{{Hash-Based Tree Similarity and Simplification in
  Genetic Programming for Symbolic Regression}}.
\newblock \bibinfo{journal}{arXiv e-prints} ,
  \bibinfo{pages}{arXiv:2107.10640}\DOIprefix\doi{10.48550/arXiv.2107.10640},
  \href{http://arxiv.org/abs/2107.10640}{{\tt arXiv:2107.10640}}.
\bibitem[{Cherry(1985)}]{Cherry1985}
\bibinfo{author}{Cherry, G.W.}, \bibinfo{year}{1985}.
\newblock \bibinfo{title}{Integration in finite terms with special functions:
  the error function}.
\newblock \bibinfo{journal}{Journal of Symbolic Computation}
  \bibinfo{volume}{1}, \bibinfo{pages}{283--302}.
\newblock \DOIprefix\doi{10.1016/S0747-7171(85)80037-7}.
\bibitem[{Cherry(1986)}]{Cherry1986}
\bibinfo{author}{Cherry, G.W.}, \bibinfo{year}{1986}.
\newblock \bibinfo{title}{Integration in finite terms with special functions:
  the logarithmic integral}.
\newblock \bibinfo{journal}{SIAM Journal on Computing} \bibinfo{volume}{15},
  \bibinfo{pages}{1--12}.
\newblock \DOIprefix\doi{10.1137/0215001}.
\bibitem[{{Desmond} et~al.(2023){Desmond}, {Bartlett} and {Ferreira}}]{RAR}
\bibinfo{author}{{Desmond}, H.}, \bibinfo{author}{{Bartlett}, D.J.},
  \bibinfo{author}{{Ferreira}, P.G.}, \bibinfo{year}{2023}.
\newblock \bibinfo{title}{{On the functional form of the radial acceleration
  relation}}.
\newblock \bibinfo{journal}{Mon. Not. Roy. Aston. Soc.} \bibinfo{volume}{521},
  \bibinfo{pages}{1817--1831}.
\newblock \DOIprefix\doi{10.1093/mnras/stad597},
  \href{http://arxiv.org/abs/2301.04368}{{\tt arXiv:2301.04368}}.
\bibitem[{DLMF()}]{DLMF}
DLMF, .
\newblock \bibinfo{title}{{NIST} digital library of mathematical functions}.
\newblock \bibinfo{howpublished}{\url{https://dlmf.nist.gov/}}.
\newblock \bibinfo{note}{F.~W.~J. Olver, A.~B. {Olde Daalhuis}, D.~W. Lozier,
  B.~I. Schneider, R.~F. Boisvert, C.~W. Clark, B.~R. Miller, B.~V. Saunders,
  H.~S. Cohl, and M.~A. McClain, eds.}
\bibitem[{Flajolet and Sedgewick(2009)}]{FlajoletSedgewick2009}
\bibinfo{author}{Flajolet, P.}, \bibinfo{author}{Sedgewick, R.},
  \bibinfo{year}{2009}.
\newblock \bibinfo{title}{Analytic Combinatorics}.
\newblock \bibinfo{publisher}{Cambridge University Press}.
\bibitem[{{Ford} et~al.(2026){Ford}, {Desmond}, {Bartlett} and
  {Ferreira}}]{Ford}
\bibinfo{author}{{Ford}, A.}, \bibinfo{author}{{Desmond}, H.},
  \bibinfo{author}{{Bartlett}, D.J.}, \bibinfo{author}{{Ferreira}, P.G.},
  \bibinfo{year}{2026}.
\newblock \bibinfo{title}{{The functional form of galaxy and halo luminosity
  and mass functions}}.
\newblock \bibinfo{journal}{arXiv e-prints} ,
  \bibinfo{pages}{arXiv:2604.23236}\DOIprefix\doi{10.48550/arXiv.2604.23236},
  \href{http://arxiv.org/abs/2604.23236}{{\tt arXiv:2604.23236}}.
\bibitem[{{FriCAS team}(2024)}]{FriCAS2024}
\bibinfo{author}{{FriCAS team}}, \bibinfo{year}{2024}.
\newblock \bibinfo{title}{{FriCAS}: an advanced computer algebra system}.
\newblock \bibinfo{note}{Version 1.3.10. Available at
  \url{https://fricas.github.io/}. Fork of Axiom with the most complete
  open-source implementation of the Risch algorithm}.
\bibitem[{Gradshteyn and Ryzhik(2015)}]{GradshteynRyzhik2015}
\bibinfo{author}{Gradshteyn, I.S.}, \bibinfo{author}{Ryzhik, I.M.},
  \bibinfo{year}{2015}.
\newblock \bibinfo{title}{Table of Integrals, Series, and Products}.
\newblock \bibinfo{edition}{8th} ed., \bibinfo{publisher}{Academic Press},
  \bibinfo{address}{Boston}.
\newblock \bibinfo{note}{Edited by D. Zwillinger and V. H. Moll}.
\bibitem[{Hebisch(2018)}]{Hebisch2018}
\bibinfo{author}{Hebisch, W.}, \bibinfo{year}{2018}.
\newblock \bibinfo{title}{Integration in terms of exponential integrals and
  incomplete gamma functions {I}}.
\newblock \bibinfo{journal}{arXiv preprint}
  \href{http://arxiv.org/abs/1802.05544}{{\tt arXiv:1802.05544}}.
\bibitem[{Kauers and Koutschan(2015)}]{KauersKoutschan2015}
\bibinfo{author}{Kauers, M.}, \bibinfo{author}{Koutschan, C.},
  \bibinfo{year}{2015}.
\newblock \bibinfo{title}{Integral {D}-finite functions}, in:
  \bibinfo{booktitle}{Proceedings of the 2015 {ACM} International Symposium on
  Symbolic and Algebraic Computation ({ISSAC} 2015)}, \bibinfo{publisher}{ACM},
  \bibinfo{address}{Bath, United Kingdom}. pp. \bibinfo{pages}{251--258}.
\newblock \DOIprefix\doi{10.1145/2755996.2756658},
  \href{http://arxiv.org/abs/1501.03691}{{\tt arXiv:1501.03691}}.
\bibitem[{Koutschan(2013)}]{Koutschan2013}
\bibinfo{author}{Koutschan, C.}, \bibinfo{year}{2013}.
\newblock \bibinfo{title}{Creative telescoping for holonomic functions}, in:
  \bibinfo{editor}{Schneider, C.}, \bibinfo{editor}{Bl\"umlein, J.} (Eds.),
  \bibinfo{booktitle}{Computer Algebra in Quantum Field Theory: Integration,
  Summation and Special Functions}. \bibinfo{publisher}{Springer Vienna}. Texts
  \& Monographs in Symbolic Computation, pp. \bibinfo{pages}{171--194}.
\newblock \DOIprefix\doi{10.1007/978-3-7091-1616-6_7},
  \href{http://arxiv.org/abs/1307.4554}{{\tt arXiv:1307.4554}}.
\bibitem[{Kronberger et~al.(2024)Kronberger, {Olivetti de Franca}, Desmond,
  Bartlett and Kammerer}]{Kronberger2024}
\bibinfo{author}{Kronberger, G.}, \bibinfo{author}{{Olivetti de Franca}, F.},
  \bibinfo{author}{Desmond, H.}, \bibinfo{author}{Bartlett, D.J.},
  \bibinfo{author}{Kammerer, L.}, \bibinfo{year}{2024}.
\newblock \bibinfo{title}{The inefficiency of genetic programming for symbolic
  regression}.
\newblock \bibinfo{journal}{Parallel Problem Solving from Nature (PPSN XVIII)}
  \bibinfo{volume}{15148}, \bibinfo{pages}{273--290}.
\newblock \href{http://arxiv.org/abs/2404.17292}{{\tt arXiv:2404.17292}}.
\bibitem[{Lample and Charton(2020)}]{LampleCharton2020}
\bibinfo{author}{Lample, G.}, \bibinfo{author}{Charton, F.},
  \bibinfo{year}{2020}.
\newblock \bibinfo{title}{Deep learning for symbolic mathematics}, in:
  \bibinfo{booktitle}{International Conference on Learning Representations
  (ICLR)}, pp. \bibinfo{pages}{1--16}.
\newblock \href{http://arxiv.org/abs/1912.01412}{{\tt arXiv:1912.01412}}.
\bibitem[{Liouville(1835)}]{Liouville1835}
\bibinfo{author}{Liouville, J.}, \bibinfo{year}{1835}.
\newblock \bibinfo{title}{{M\'emoire sur l'int\'egration d'une classe de
  fonctions transcendantes}}.
\newblock \bibinfo{journal}{Journal f\"ur die reine und angewandte Mathematik}
  \bibinfo{volume}{13}, \bibinfo{pages}{93--118}.
\newblock \DOIprefix\doi{10.1515/crll.1835.13.93}.
\bibitem[{Magid(1994)}]{Magid1994}
\bibinfo{author}{Magid, A.R.}, \bibinfo{year}{1994}.
\newblock \bibinfo{title}{Lectures on Differential {G}alois Theory}.
  volume~\bibinfo{volume}{7} of \textit{\bibinfo{series}{University Lecture
  Series}}.
\newblock \bibinfo{publisher}{American Mathematical Society},
  \bibinfo{address}{Providence, RI}.
\bibitem[{{Mart{\'\i}n} et~al.(2025){Mart{\'\i}n}, {Yasin}, {Bartlett},
  {Desmond} and {Ferreira}}]{Martin_1}
\bibinfo{author}{{Mart{\'\i}n}, A.}, \bibinfo{author}{{Yasin}, T.},
  \bibinfo{author}{{Bartlett}, D.J.}, \bibinfo{author}{{Desmond}, H.},
  \bibinfo{author}{{Ferreira}, P.G.}, \bibinfo{year}{2025}.
\newblock \bibinfo{title}{{Constraining dark matter halo profiles with symbolic
  regression}}.
\newblock \bibinfo{journal}{arXiv e-prints} ,
  \bibinfo{pages}{arXiv:2511.23073}\DOIprefix\doi{10.48550/arXiv.2511.23073},
  \href{http://arxiv.org/abs/2511.23073}{{\tt arXiv:2511.23073}}.
\bibitem[{{Mart{\'\i}n} et~al.(2026){Mart{\'\i}n}, {Yasin}, {Bartlett},
  {Desmond} and {Ferreira}}]{Martin_2}
\bibinfo{author}{{Mart{\'\i}n}, A.}, \bibinfo{author}{{Yasin}, T.},
  \bibinfo{author}{{Bartlett}, D.J.}, \bibinfo{author}{{Desmond}, H.},
  \bibinfo{author}{{Ferreira}, P.G.}, \bibinfo{year}{2026}.
\newblock \bibinfo{title}{{Symbolically regressing dark matter halo profiles
  using weak lensing}}.
\newblock \bibinfo{journal}{arXiv e-prints} ,
  \bibinfo{pages}{arXiv:2601.05203}\DOIprefix\doi{10.48550/arXiv.2601.05203},
  \href{http://arxiv.org/abs/2601.05203}{{\tt arXiv:2601.05203}}.
\bibitem[{{Maxima developers}(2024)}]{Maxima2024}
\bibinfo{author}{{Maxima developers}}, \bibinfo{year}{2024}.
\newblock \bibinfo{title}{Maxima, a computer algebra system}.
\newblock \bibinfo{note}{Version 5.47. Available at
  \url{https://maxima.sourceforge.io/}}.
\bibitem[{Meurer et~al.(2017)Meurer, Smith, Paprocki, \v{C}ert\'{i}k,
  Kirpichev, Rocklin, Kumar, Ivanov, Moore, Singh, Rathnayake, Vig, Granger,
  Muller, Bonazzi, Gupta, Vats, Johansson, Pedregosa, Curry, Terrel,
  \v{S}t\v{e}p\'{a}n Rou\v{c}ka, Saboo, Fernando, Kulal, Cimrman and
  Scopatz}]{Meurer2017}
\bibinfo{author}{Meurer, A.}, \bibinfo{author}{Smith, C.P.},
  \bibinfo{author}{Paprocki, M.}, \bibinfo{author}{\v{C}ert\'{i}k, O.},
  \bibinfo{author}{Kirpichev, S.B.}, \bibinfo{author}{Rocklin, M.},
  \bibinfo{author}{Kumar, A.}, \bibinfo{author}{Ivanov, S.},
  \bibinfo{author}{Moore, J.K.}, \bibinfo{author}{Singh, S.},
  \bibinfo{author}{Rathnayake, T.}, \bibinfo{author}{Vig, S.},
  \bibinfo{author}{Granger, B.E.}, \bibinfo{author}{Muller, R.P.},
  \bibinfo{author}{Bonazzi, F.}, \bibinfo{author}{Gupta, H.},
  \bibinfo{author}{Vats, S.}, \bibinfo{author}{Johansson, F.},
  \bibinfo{author}{Pedregosa, F.}, \bibinfo{author}{Curry, M.J.},
  \bibinfo{author}{Terrel, A.R.}, \bibinfo{author}{\v{S}t\v{e}p\'{a}n
  Rou\v{c}ka}, \bibinfo{author}{Saboo, A.}, \bibinfo{author}{Fernando, I.},
  \bibinfo{author}{Kulal, S.}, \bibinfo{author}{Cimrman, R.},
  \bibinfo{author}{Scopatz, A.}, \bibinfo{year}{2017}.
\newblock \bibinfo{title}{{SymPy}: Symbolic computing in {Python}}.
\newblock \bibinfo{journal}{PeerJ Computer Science} \bibinfo{volume}{3},
  \bibinfo{pages}{e103}.
\newblock \DOIprefix\doi{10.7717/peerj-cs.103}.
\bibitem[{Moses(1971)}]{Moses1971}
\bibinfo{author}{Moses, J.}, \bibinfo{year}{1971}.
\newblock \bibinfo{title}{Symbolic integration: The stormy decade}.
\newblock \bibinfo{journal}{Communications of the ACM} \bibinfo{volume}{14},
  \bibinfo{pages}{548--560}.
\newblock \DOIprefix\doi{10.1145/362637.362651}.
\bibitem[{Norman and Moore(1977)}]{NormanMoore1977}
\bibinfo{author}{Norman, A.C.}, \bibinfo{author}{Moore, P.M.A.},
  \bibinfo{year}{1977}.
\newblock \bibinfo{title}{Implementing the new {R}isch integration algorithm},
  in: \bibinfo{booktitle}{Proceedings of the 4th International Colloquium on
  Advanced Computing Methods in Theoretical Physics},
  \bibinfo{address}{Marseilles, France}. pp. \bibinfo{pages}{99--110}.
\bibitem[{Parisse and {De Graeve}(2024)}]{Giac2024}
\bibinfo{author}{Parisse, B.}, \bibinfo{author}{{De Graeve}, R.},
  \bibinfo{year}{2024}.
\newblock \bibinfo{title}{Giac/xcas, a free computer algebra system}.
\newblock \bibinfo{note}{Version 1.9. Available at
  \url{https://www-fourier.ujf-grenoble.fr/~parisse/giac.html}}.
\bibitem[{Prudnikov et~al.(1986)Prudnikov, Brychkov and
  Marichev}]{Prudnikov1986}
\bibinfo{author}{Prudnikov, A.P.}, \bibinfo{author}{Brychkov, Y.A.},
  \bibinfo{author}{Marichev, O.I.}, \bibinfo{year}{1986}.
\newblock \bibinfo{title}{Integrals and Series}. volume \bibinfo{volume}{1:
  Elementary Functions}.
\newblock \bibinfo{publisher}{Gordon and Breach Science Publishers},
  \bibinfo{address}{New York}.
\bibitem[{Raab(2012)}]{Raab2012}
\bibinfo{author}{Raab, C.G.}, \bibinfo{year}{2012}.
\newblock \bibinfo{title}{Definite Integration in Differential Fields}.
\newblock Ph.D. thesis. Johannes Kepler Universit\"at Linz.
  \bibinfo{address}{Linz, Austria}.
\newblock \bibinfo{note}{RISC Doctoral Program Computational Mathematics}.
\bibitem[{Raab(2013)}]{Raab2013}
\bibinfo{author}{Raab, C.G.}, \bibinfo{year}{2013}.
\newblock \bibinfo{title}{Generalization of {R}isch's algorithm to special
  functions}, in: \bibinfo{editor}{Schneider, C.}, \bibinfo{editor}{Bl\"umlein,
  J.} (Eds.), \bibinfo{booktitle}{Computer Algebra in Quantum Field Theory:
  Integration, Summation and Special Functions}. \bibinfo{publisher}{Springer
  Vienna}. Texts \& Monographs in Symbolic Computation, pp.
  \bibinfo{pages}{285--304}.
\newblock \DOIprefix\doi{10.1007/978-3-7091-1616-6_12},
  \href{http://arxiv.org/abs/1305.1481}{{\tt arXiv:1305.1481}}.
\bibitem[{Rich et~al.(2018)Rich, Scheibe and Abbasi}]{RichScheibeAbbasi2018}
\bibinfo{author}{Rich, A.}, \bibinfo{author}{Scheibe, P.},
  \bibinfo{author}{Abbasi, N.M.}, \bibinfo{year}{2018}.
\newblock \bibinfo{title}{Rule-based integration: An extensive system of
  symbolic integration rules}.
\newblock \bibinfo{journal}{Journal of Open Source Software}
  \bibinfo{volume}{3}, \bibinfo{pages}{1073}.
\newblock \DOIprefix\doi{10.21105/joss.01073}.
\bibitem[{Rich(2024)}]{Rubi2024}
\bibinfo{author}{Rich, A.D.}, \bibinfo{year}{2024}.
\newblock \bibinfo{title}{{Rubi}: Rule-based integration}.
\newblock \bibinfo{note}{Version 4.17.3. Available at
  \url{https://rulebasedintegration.org/}. Approximately 6,700 integration
  rules implemented in Mathematica}.
\bibitem[{Risch(1969)}]{Risch1969}
\bibinfo{author}{Risch, R.H.}, \bibinfo{year}{1969}.
\newblock \bibinfo{title}{The problem of integration in finite terms}.
\newblock \bibinfo{journal}{Transactions of the American Mathematical Society}
  \bibinfo{volume}{139}, \bibinfo{pages}{167--189}.
\newblock \DOIprefix\doi{10.1090/S0002-9947-1969-0237477-8}.
\bibitem[{Rosenlicht(1972)}]{Rosenlicht1972}
\bibinfo{author}{Rosenlicht, M.}, \bibinfo{year}{1972}.
\newblock \bibinfo{title}{Integration in finite terms}.
\newblock \bibinfo{journal}{The American Mathematical Monthly}
  \bibinfo{volume}{79}, \bibinfo{pages}{963--972}.
\newblock \DOIprefix\doi{10.1080/00029890.1972.11993166}.
\bibitem[{{Sousa} et~al.(2024){Sousa}, {Bartlett}, {Desmond} and
  {Ferreira}}]{Inflation}
\bibinfo{author}{{Sousa}, T.}, \bibinfo{author}{{Bartlett}, D.J.},
  \bibinfo{author}{{Desmond}, H.}, \bibinfo{author}{{Ferreira}, P.G.},
  \bibinfo{year}{2024}.
\newblock \bibinfo{title}{{Optimal inflationary potentials}}.
\newblock \bibinfo{journal}{Phys. Rev. D} \bibinfo{volume}{109},
  \bibinfo{pages}{083524}.
\newblock \DOIprefix\doi{10.1103/PhysRevD.109.083524},
  \href{http://arxiv.org/abs/2310.16786}{{\tt arXiv:2310.16786}}.
\bibitem[{Trager(1976)}]{Trager1976}
\bibinfo{author}{Trager, B.M.}, \bibinfo{year}{1976}.
\newblock \bibinfo{title}{Algebraic Factoring and Rational Function
  Integration}.
\newblock Ph.D. thesis. Massachusetts Institute of Technology.
\newblock \bibinfo{note}{S.M. thesis}.
\bibitem[{{Wolfram Research, Inc.}(2024)}]{Mathematica}
\bibinfo{author}{{Wolfram Research, Inc.}}, \bibinfo{year}{2024}.
\newblock \bibinfo{title}{Mathematica, {V}ersion 14.0}.
\newblock \URLprefix \url{https://www.wolfram.com/mathematica}.
  \bibinfo{note}{champaign, IL}.

\end{thebibliography}

\end{document}